# Unified treatment of spin torques using a coupled magnetisation dynamics and three-dimensional spin current solver

Serban Lepadatu[*]

*Jeremiah Horrocks Institute for Mathematics, Physics and Astronomy, University of Central Lancashire, Preston PR1 2HE, U.K.*

Abstract

A three-dimensional spin current solver based on a generalised spin drift-diffusion description, including the spin Hall effect, is integrated with a magnetisation dynamics solver. The resulting model is shown to simultaneously reproduce the spin-orbit torques generated using the spin Hall effect, spin pumping torques generated by magnetisation dynamics in multilayers, as well as the spin transfer torques acting on magnetisation regions with spatial gradients, whilst field-like and spin-like torques are reproduced in a spin valve geometry. Two approaches to modelling interfaces are analysed, one based on the spin mixing conductance and the other based on continuity of spin currents where the spin dephasing length governs the absorption of transverse spin components. In both cases analytical formulas are derived for the spin-orbit torques in a heavy metal / ferromagnet bilayer geometry, showing in general both field-like and damping-like torques are generated. The limitations of the analytical approach are discussed, showing that even in a simple bilayer geometry, due to the non-uniformity of the spin currents, a full three-dimensional treatment is required. Finally the model is applied to the quantitative analysis of the spin Hall angle in Pt by reproducing published experimental data on the ferromagnetic resonance linewidth in the bilayer geometry.

[*] SLepadatu@uclan.ac.uk

# I. Introduction

The study of spin torques is currently of great interest due to applications to magnetic devices, including race-track memory [1] and magnetic tunnel junction devices for memory and spin logic applications [2]. Spin transport in magnetic multilayers is a complex and intensely researched topic, with many sources of spin torques identified, ranging from bulk spin transport phenomena including the spin Hall effect (SHE) [3], to interfacial Rashba-type spin transfer phenomena [4] including inverse spin galvanic effects [5] and intrinsic spin-orbit torques originating from the Berry phase [6]. The usual approach to analysing the effect of various types of spin torques on the magnetisation structure and dynamics is based on introducing separate analytical formulations for the spin torques within a magnetisation dynamics model. As devices become more complex, involving several layers and multiple sources of spin torques, more advanced approaches are required capable of resolving the non-local nature of spin transport and non-uniformity of spin currents, which lead to a complex spatial dependence of the spin torques that cannot be fully accounted for within an analytical formulation. An alternative approach describes the flow of charges and spins using a drift-diffusion model [7-10]. The incorporation of a drift-diffusion formulation within a micromagnetics model is currently of great interest [11-16]. An important source of spin currents which needs to be included within a three-dimensional model of magnetisation dynamics is the SHE. This effect was first predicted by Dyakonov and Perel [3], where due to spin-orbit interaction [17] an electrical current results in transverse flow of spins with polarisation perpendicular to both the charge and spin current directions [18]. The SHE was demonstrated using a number of techniques, including field-swept ferromagnetic resonance (FMR) [19,20], spin torque FMR [21-23], optical FMR [24], time-resolved optical techniques [25] and electrical methods [26]. The SHE-generated spin polarisation results in a torque when injected in a ferromagnetic layer [27,28], which makes this effect very important for magnetic devices, resulting in motion of domain walls [29] with important applications to synthetic antiferromagnetic domain wall devices [30,31]. The reverse effect also exists where a spin current, typically generated through spin pumping from a ferromagnetic layer, results in a transverse displacement of charges through the inverse SHE [32-34]. An important parameter that characterizes both the SHE and its inverse is the spin Hall angle. This has been measured in various heavy metals – for reviews see Refs. [35,36] – but remains a topic of contention; in particular for Pt there is a large disparity in reported results, spanning a range of around 30 times from ~0.004 to over 0.1. The spin pumping effect itself also generates a spin torque on a dynamically excited magnetisation texture due to loss of spin angular momentum [37].

Here the SHE and spin pumping effect are incorporated within a three-dimensional generalised drift-diffusion model, and coupled to a magnetisation dynamics solver. The implementation of the model is presented in Section II and shown to self-consistently reproduce a number of spin torques within the same description, namely the SHE-generated spin-orbit torques (SOT), torques due to loss of spin angular momentum via spin pumping, spin transfer torques (STT) [38] arising in the presence of magnetisation structures with spatial gradients, as well as the field-like and Slonczewski spin torques in spin valve geometries [39,40]. The SOT in a bilayer geometry is analysed in some detail in Section III, obtaining analytical expressions for two approaches to modelling spin transfer between layers: one based on the spin mixing conductance at the interface [41], and the other based on continuity of spin currents, with the absorption of transverse spin components governed by the spin dephasing length; the limitations of the analytical approach are also discussed. Finally the model is applied to the analysis of FMR linewidth in Section IV, obtaining estimations of the spin Hall angle in Pt based on both the spin mixing conductance and spin dephasing length absorption approaches.

## II. Spin Drift-Diffusion Model and Implementation

The flow of charges and spins in a multi-layered structure can be described as [18,7]:

$$\mathbf{J}_C = \sigma\mathbf{E} + \beta_D D_e \frac{e}{\mu_B}(\nabla\mathbf{S})\mathbf{m} + \theta_{SHA} D_e \frac{e}{\mu_B}\nabla\times\mathbf{S} + D_e\nabla n \tag{1}$$

$$\mathbf{J}_S = -\frac{\mu_B}{e}(\beta_\sigma \sigma\mathbf{E} + \beta_D D_e \nabla n)\otimes\mathbf{m} - D_e\nabla\mathbf{S} + \theta_{SHA}\frac{\mu_B}{e}\varepsilon(\sigma\mathbf{E} + D_e\nabla n) \tag{2}$$

Here the convention used by Dyakonov [18] has been adopted, where $\mathbf{J}_S$ is a rank-2 tensor such that $\mathbf{J}_{Sij}$ signifies the flow of the $j$ component of spin polarisation in the direction $i$.. $\mathbf{J}_C$ is the usual electrical current density and $\mathbf{J}_S$ is the spin polarisation current density, but will be improperly referred to as the spin current for brevity; $\mathbf{J}_S$ can be converted to spin current by multiplication with $\hbar/2\mu_B$. Equation (1) contains the usual Ohm's law term, where $\sigma$ is the electrical conductivity, as well as a term due to the giant magneto-resistance contribution arising in current perpendicular to the plane (CPP-GMR) stacks [7,8], where $D_e$ is the electron diffusion constant, $\beta_D$ is the diffusion spin polarisation, $\mathbf{m}$ is the magnetisation direction and $\mathbf{S}$ is the spin accumulation. The last two terms in equation (1) are the inverse spin Hall effect, where $\theta_{SHA}$ is the spin Hall angle, and diffusion of charges due to charge density gradients, where $n$ is the volume charge density; note $\sigma = \mu n$, where $\mu$ is the electron mobility. Equation (2) contains three spin current contributions: i) the flow of spins carried by a spin-polarised charge current in a magnetic layer, either due to an external electric field $\mathbf{E}$ or diffusion of charges, where $\beta_\sigma$ is the charge current spin polarisation, ii) diffusion of spins due to local spin accumulation gradients, and iii) spin current generated by the SHE, where $\boldsymbol{\varepsilon}$ is the rank-3 unit anti-symmetric tensor. Due to the complexity of the problem the implementation of equations (1) and (2) is split into two parts: here we concentrate only on the direct SHE which is responsible for generating spin torques on a ferromagnetic layer. The implementation of the inverse SHE will be addressed in a forthcoming publication; without this term charge density gradients may be ignored and equation (1) is curl-free, thus the usual relations $\nabla \cdot \mathbf{J}_c = 0$ and $\mathbf{E} = -\nabla V$ apply, with $V$ being the electrical potential.

The exchange interaction between the spin accumulation and local magnetic moment results in a torque on the magnetisation. Only the transverse component of the spin accumulation (meaning transverse to the magnetisation direction) generates a torque. In order to conserve total spin angular momentum the transverse spin components are quickly absorbed, thus the relaxation of the longitudinal and transverse spin accumulations are governed by different length scales. The decay of the longitudinal spin accumulation is a diffusive process governed by the spin-flip length $\lambda_{sf}$, whilst the relaxation of the transverse spin accumulation is a ballistic process occurring on a much shorter length scale. One of these length scales is the exchange rotation length $\lambda_J = \sqrt{\hbar D_e / J}$, where $J$ is the exchange interaction energy strength [7,8]. Another important length scale is the spin dephasing length which governs the decay of transverse spin accumulation components. This is given by $\lambda_\varphi = \lambda_J \sqrt{l_\perp / l_L}$, where $l_\perp$ and $l_L$ are the spin coherence and spin precession lengths respectively [9]. The equation of motion for spin accumulation is now given by:

$$\frac{\partial \mathbf{S}}{\partial t} = -\nabla \cdot \mathbf{J}_S - D_e \left( \frac{\mathbf{S}}{\lambda_{sf}^2} + \frac{\mathbf{S} \times \mathbf{m}}{\lambda_J^2} + \frac{\mathbf{m} \times (\mathbf{S} \times \mathbf{m})}{\lambda_\varphi^2} \right) \qquad (3)$$

The torque on the magnetisation is obtained using the same arguments given in Ref. [8], by considering the conservation of total spin angular momentum; thus in the steady state where $\partial \mathbf{S}/\partial t = 0$, this is obtained from the divergence of the spin current as:

$$\mathbf{T_s} = -\frac{D_e}{\lambda_J^2}\mathbf{m}\times\mathbf{S} - \frac{D_e}{\lambda_\varphi^2}\mathbf{m}\times(\mathbf{m}\times\mathbf{S}) \qquad (4)$$

The equation of motion for magnetisation is now a modified Landau-Lifshitz-Gilbert (LLG) equation containing the additional total spin torque as:

$$\frac{\partial \mathbf{m}}{\partial t} = -\gamma\mathbf{m}\times\mathbf{H}_{eff} + \alpha\mathbf{m}\times\frac{\partial \mathbf{m}}{\partial t} + \frac{1}{M_S}\mathbf{T_s} \qquad (5)$$

Here $\gamma = \mu_0|\gamma_e|$, where $\gamma_e = -g\mu_B/\hbar$ is the electron gyromagnetic ratio, $M_s$ is the saturation magnetisation, and $\mathbf{H}_{eff}$ contains all the usual effective field contributions, typically including demagnetising, direct exchange and applied field contributions.

The response time-scales of **m** and **S** are separated typically by 3 orders of magnitude (ps vs fs time-scales respectively) thus equations (3) and (5) may be evaluated separately. Explicitly, the following computational procedure has been implemented in the finite-difference micromagnetics-oriented software Boris [42,43]. Using the relations $\nabla.\mathbf{J}_c = 0$ and $\mathbf{E} = -\nabla V$, the following Poisson equation is obtained from equation (1):

$$\nabla^2 V = -\frac{(\nabla V)(\nabla \sigma)}{\sigma} + \frac{\beta_D D_e}{\sigma}\frac{e}{\mu_B}\nabla.(\nabla\mathbf{S})\mathbf{m} \qquad (6)$$

In general the conductivity is allowed to vary within the same material even for uniform charge density, for example due to anisotropic MR in magnetic layers [43], however it is not included in this work. Equation (6) is evaluated for a given spin accumulation and fixed potential boundary conditions on two electrodes. Using the calculated electrical potential, the charge current density is obtained using equation (1) and substituted in equation (2) to obtain the spin current density as:

$$\mathbf{J}_S = -\beta_\sigma \frac{\mu_B}{e} \mathbf{J}_C \otimes \mathbf{m} + \beta_\sigma \beta_\mathbf{D} D_e [(\nabla \mathbf{S})\mathbf{m}] \otimes \mathbf{m} - D_e \nabla \mathbf{S} + \theta_{SHA} \frac{\mu_B}{e} \varepsilon \mathbf{J}_C \qquad (7)$$

Finally from equations (3) and (7) the spin accumulation equation of motion is rewritten in terms of the charge current density as:

$$\begin{aligned}\frac{\partial \mathbf{S}}{\partial t} &= \beta_\sigma \frac{\mu_B}{e}(\mathbf{J}_C.\nabla)\mathbf{m} - \beta_\sigma \beta_\mathbf{D} D_e \{[(\nabla \mathbf{S})\mathbf{m}.\nabla]\mathbf{m} + \mathbf{m}\nabla.(\nabla \mathbf{S})\mathbf{m}\} - \theta_{SHA} \frac{\mu_B}{e} \nabla.(\varepsilon \mathbf{J}_C) \\ &+ D_e \nabla^2 \mathbf{S} - D_e \left( \frac{\mathbf{S}}{\lambda_{sf}^2} + \frac{\mathbf{S}\times\mathbf{m}}{\lambda_J^2} + \frac{\mathbf{m}\times(\mathbf{S}\times\mathbf{m})}{\lambda_\varphi^2} \right) \end{aligned} \qquad (8)$$

Equation (8) is solved to obtain the steady state spin accumulation by setting $\partial \mathbf{S}/\partial t = 0$. For the spatial discretization a multi-level multi-grid method is used [44], with equation (5) evaluated on a coarse mesh, whilst equations (6) and (8) are evaluated on a sufficiently refined sub-mesh; all meshes use rectangular prism cells with the *z* cellsize independent of the *xy*-plane cellsize.

For a multi-layered geometry it is important to consider both the interface and mesh boundary conditions. Boundary conditions for evaluating differential operators are derived from the physically motivated requirements that both the charge and spin currents perpendicular to a mesh boundary not containing an electrode are zero: $\mathbf{J}_C.\mathbf{n} = 0$ and $\mathbf{J}_S.\mathbf{n} = 0$ [17], where **n** is the boundary normal. In this case we obtain the following Neumann boundary conditions from equations (1) and (2):

$$\begin{aligned} \nabla V.\mathbf{n} &= 0 \\ (\nabla \mathbf{S})\mathbf{n} &= \frac{\theta_{SHA}}{D_e} \frac{\mu_B}{e}(\varepsilon \mathbf{J}_C)\mathbf{n} \end{aligned} \qquad (9)$$

For boundaries containing an electrode *V* is specified on the boundary thus $\nabla V.\mathbf{n}$ is also prescribed. The spin current perpendicular to an electrode is not zero in general and electrode-containing boundaries need special consideration to ensure physically valid results.

One general principle is to define the problem geometry and electrical contacts such that the spin accumulation gradients normal to the electrodes are zero (in particular it may be necessary to allow the spin accumulation itself to decay to zero by including in the model a sufficiently large part of the electrical contacts) – again we can use $(\nabla \mathbf{S}).\mathbf{n} = 0$; for magnetic regions this further requires the magnetisation be uniform around the electrodes.

In the transparent interface limit where specular scattering can be neglected, values of $V$ and $\mathbf{S}$ at the interface cells can be derived by enforcing the continuity of both $\mathbf{J}_C.\mathbf{n}$ and $\mathbf{J}_S.\mathbf{n}$ [8,45]. In this picture the absorption of transverse spin components is governed by the length-scales $\lambda_J$ and $\lambda_\varphi$. An alternative approach is that of magnetoelectric circuit theory [41], where the absorption of transverse spin components is confined to the interface and modelled via the complex spin mixing conductance $G^{\uparrow\downarrow}$. The boundary conditions for the charge and spin currents at a normal metal (N) / ferromagnet (F) interface are written as:

$$\mathbf{J}_C.\mathbf{n}\big|_N = \mathbf{J}_C.\mathbf{n}\big|_F = -(G^\uparrow + G^\downarrow)\Delta V + (G^\uparrow - G^\downarrow)\Delta \mathbf{V}_S.\mathbf{m}$$

$$\mathbf{J}_S.\mathbf{n}\big|_N - \mathbf{J}_S.\mathbf{n}\big|_F = \frac{2\mu_B}{e}\left[\mathrm{Re}\{G^{\uparrow\downarrow}\}\mathbf{m}\times(\mathbf{m}\times\Delta\mathbf{V}_S) + \mathrm{Im}\{G^{\uparrow\downarrow}\}\mathbf{m}\times\Delta\mathbf{V}_S\right] \quad (10)$$

$$\mathbf{J}_S.\mathbf{n}\big|_F = \frac{\mu_B}{e}\left[-(G^\uparrow + G^\downarrow)(\Delta\mathbf{V}_S.\mathbf{m})\mathbf{m} + (G^\uparrow - G^\downarrow)\Delta V\mathbf{m}\right]$$

Here $\Delta V$ is the potential drop across the N/F interface ($\Delta V = V_F - V_N$) and $\Delta \mathbf{V}_S$ is the spin chemical potential drop, where $\mathbf{V}_S = (D_e/\sigma)(e/\mu_B)\mathbf{S}$, and $G^\uparrow$, $G^\downarrow$ are interface conductances for the majority and minority spin carriers respectively. Equation (10) together with equations (1) and (7) are used to calculate the potential and spin accumulation either side of the boundary; the transverse spin current absorbed at the interface then gives rise to a torque which may be included in the magnetic cells at the interface with cellsize $d_h$, in addition to any other torques resulting from transverse spin accumulation in Eq. (4), as:

$$\mathbf{T}_S^{\text{interface}} = \frac{g\mu_B}{ed_h}\left[\text{Re}\{G^{\uparrow\downarrow}\}\mathbf{m}\times(\mathbf{m}\times\Delta\mathbf{V}_S) + \text{Im}\{G^{\uparrow\downarrow}\}\mathbf{m}\times\Delta\mathbf{V}_S\right] \quad (11)$$

The Onsager reciprocal process to absorption of transverse spin currents is the generation of spin currents via dynamical magnetisation processes, e.g. magnetisation precession, known as spin pumping [37,46]. This is given in equation (12), where $g^{\uparrow\downarrow} = (h/e^2)G^{\uparrow\downarrow}$, and may be included on the N side of equation (10) when calculating the spin chemical potential drop.

$$\mathbf{J}_S^{pump} = \frac{\mu_B}{2\pi}\left[\text{Re}\{g^{\uparrow\downarrow}\}\mathbf{m}\times\frac{\partial\mathbf{m}}{\partial t} + \text{Im}\{g^{\uparrow\downarrow}\}\frac{\partial\mathbf{m}}{\partial t}\right] \quad (12)$$

The implemented model is now applied to a N/F bilayer geometry, similar to that used in FMR experiments [20,21]. The diffusion spin polarisation, $\beta_D$, is set to zero for this geometry – for completeness the effect of spin torques in a CPP-GMR stack is addressed in Appendix A. For the N and F layers parameters associated with Pt and $Ni_{80}Fe_{20}$ (Py) are used as in Ref. [20]; $\theta_{SHA}$ is initially set to 0.1. Additionally for metals $D_e = 10^{-2}$ m$^2$/s [47]; from equation (4) it appears the spin torque is proportional to $D_e$, however the spin accumulation is inversely proportional to $D_e$, thus if the two metal layers have similar diffusion constants the spin torque is independent of $D_e$. Using a non-adiabaticity parameter $\xi = 0.04$ for Py [48], $\lambda_J \cong 0.8$ nm is determined using the relation $\xi = \lambda_J^2/\lambda_{sf}^2$, and $\lambda_\varphi = \lambda_J\sqrt{l_\perp/l_L} \cong 0.6$ nm is further obtained by using the values $l_\perp = 0.9$ nm and $l_L = 1.4$ nm from Ref. [9]. The problem geometry is shown in Figure 1.

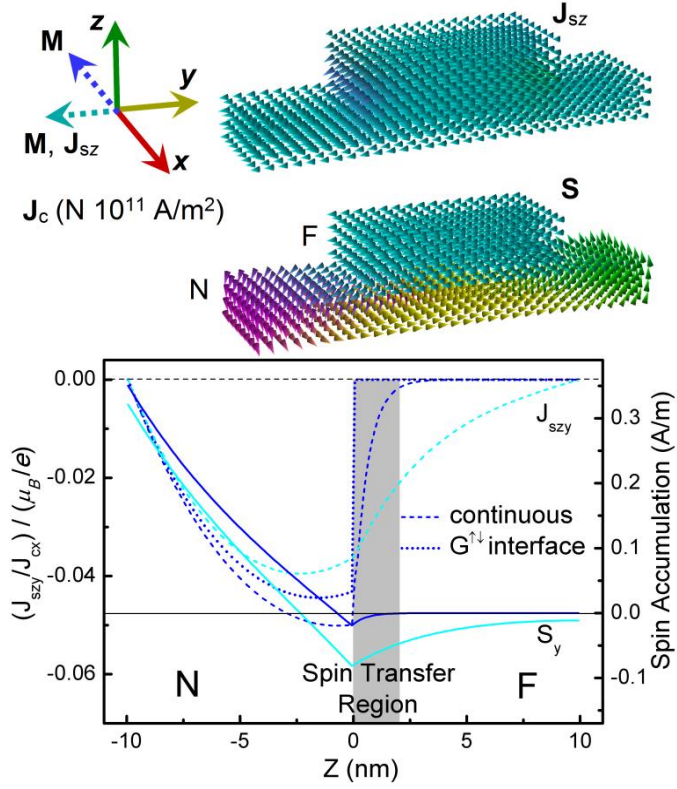

**Figure 1. Spin accumulation and spin currents in a bilayer geometry.** The bilayer consists of N (10 nm) / F (10 nm) layers with electrical contacts at the *x*-axis ends of the geometry. The computed spin accumulation **S** and perpendicular spin current $\mathbf{J}_{sz}$ are rendered for uniform magnetisation along the -*y* direction. The width of the N layer is 160 nm (the rendered plots use different display scaling factors along the *z* and *xy*-plane directions respectively for clarity). The decay of the *y* components of **S** and $\mathbf{J}_{sz}$ are shown for the longitudinal mode (magnetisation along -*y*) and transverse mode (magnetisation along -*x*). The transverse components decay on a much shorter length scale governed by the spin dephasing length, which defines a narrow spin transfer region where the spin torque is exerted on the magnetisation. For comparison, the transverse spin current is also shown for an interface where the absorption of transverse spin components is modelled via the spin mixing conductance $G^{\uparrow\downarrow}$.

As an initial test the magnetisation is set along the –*y* direction so the spin torque on the F layer is zero. As shown in Figure 1, the spin accumulation in the N layer follows the right-hand rule around the charge current direction and decays in the F layer as governed by $\lambda_{sf}^{F}$. The SHE-generated spin polarisation is perpendicular to both the charge and spin current directions and decays when injected in the F layer, reaching zero at the edges as expected. Equations (3) and (7) may be solved analytically to obtain the following expression for the value of spin polarisation at the interface in the longitudinal configuration, where $d_N$ and $d_F$ are the thickness values of the N and F layers respectively, and the quantities $N_\lambda$ and $F_\lambda$ are defined as $N_\lambda = tanh(d_N/\lambda_{sf}^N)/\lambda_{sf}^N$, $F_\lambda = tanh(d_D/\lambda_{sf}^F)/\lambda_{sf}^F$:

$$J_{szy} = -\theta_{SHA} \frac{\mu_B}{e} J_{cx} \left(1 - \frac{1}{cosh(d_N/\lambda_{sf}^N)}\right) \bigg/ \left(1 + \frac{N_\lambda}{F_\lambda}\right) \qquad (13)$$

III. Spin Torques

When the magnetisation is not aligned with the *y*-axis, due to conservation of total spin angular momentum, the absorption of transverse spin components results in a spin torque. For interfaces modelled using magnetoelectric circuit theory the absorption occurs purely at the interface resulting in the interfacial torque of equation (11). This sets any transverse spin components to zero on the F side. For interfaces where the spin current is continuous, the generation of the spin torque in equation (4) is accompanied by a rapid, but gradual, absorption of transverse spin components. These cases are exemplified in Figure 1 where, for the transverse configuration in the continuous case, the injected spin polarisation rapidly decays within a spin transfer region, dependent principally on $\lambda_\varphi$. For thin F layers the direct exchange interaction acts to keep the magnetisation constant along the *z* direction and thus in both cases we may define locally in the *xy*-plane a net, or average, spin torque on the F layer. First the continuous interface case is analysed. It is known that both field-like (FL) and (DL) torques act on the F layer [28]. The total spin torque may be decomposed into these components as shown in equation (14), where <**S**> is the spin accumulation averaged along the z direction, $\boldsymbol{\sigma} = \hat{\mathbf{x}} \times \hat{\mathbf{z}} = -\hat{\mathbf{y}}$, $f$ is a factor dependent on the various length scales, and $r$ is the ratio of the FL and DL torque magnitudes.

$$\mathbf{T_S} = -\frac{D_e}{\lambda_J^2}\mathbf{m}\times\langle\mathbf{S}\rangle - \frac{D_e}{\lambda_\varphi^2}\mathbf{m}\times(\mathbf{m}\times\langle\mathbf{S}\rangle) = \frac{fJ_{szy}}{d_F}[\mathbf{m}\times(\mathbf{m}\times\boldsymbol{\sigma}) + r\mathbf{m}\times\boldsymbol{\sigma}] = \mathbf{T_{DL}} + \mathbf{T_{FL}} \quad (14)$$

Expressions for $f$ and $r$ may be derived in this bilayer geometry by solving equations (3) and (7) in the transverse configuration and equating coefficients in equation (14). By introducing the length scales $\lambda_\perp$, $\lambda$, $\lambda_+$, $\lambda_-$ defined by the relations $1/\lambda_\perp^2 = 1/\lambda_{sf}^{F2} + 1/\lambda_\varphi^2$, $1/\lambda^4 = 1/\lambda_\perp^4 + 1/\lambda_J^4$, $1/\lambda_+^2 = 1/\lambda^2 + 1/\lambda_\perp^2$, $1/\lambda_-^2 = 1/\lambda^2 - 1/\lambda_\perp^2$ and further defining $C_\lambda = (1/\lambda_+ + i/\lambda_-)\tanh(d_F(1/\lambda_+ + i/\lambda_-)/\sqrt{2})/\sqrt{2}$, we obtain:

$$f = \left(1 + \frac{N_\lambda}{F_\lambda}\right)\left(\frac{Re\{G\}}{\lambda_\varphi^2} - \frac{Im\{G\}}{\lambda_J^2}\right),$$

$$r = \left(\frac{Re\{G\}}{\lambda_J^2} + \frac{Im\{G\}}{\lambda_\varphi^2}\right) \bigg/ \left(\frac{Re\{G\}}{\lambda_\varphi^2} - \frac{Im\{G\}}{\lambda_J^2}\right), \quad (15)$$

where $G = \dfrac{N_\lambda C_\lambda + |C_\lambda|^2}{(N_\lambda + Re\{C_\lambda\})^2 + Im\{C_\lambda\}^2} \dfrac{1}{(1/\lambda_\perp^2 + i/\lambda_J^2)}$

For example with the values given above for the N and F layers, $f = 1.36$ and $r = 0.03$. In a simpler analytical formulation, as used e.g. in Ref. [20], a DL torque is obtained from ballistic spin transfer at the interface [39] as $\mathbf{T}_{DL} = (J_{szy}/d_F)\mathbf{m}\times(\mathbf{m}\times\boldsymbol{\sigma})$. Note that this torque is similar to the DL torque of equation (14). Another approach to calculating interfacial spin torques is through the spin mixing conductance as in equation (11). Using the boundary conditions of equation (10), the spin torque acting on the F layer is now obtained as:

$$\mathbf{T_S} = \frac{f_G J_{szy}}{d_F}[\mathbf{m}\times(\mathbf{m}\times\boldsymbol{\sigma}) + r_G \mathbf{m}\times\boldsymbol{\sigma}],$$

where $f_G = \left(1 + \dfrac{N_\lambda}{F_\lambda}\right) \dfrac{N_\lambda Re\{\tilde{G}\} + |\tilde{G}|^2}{(N_\lambda + Re\{\tilde{G}\})^2 + Im\{\tilde{G}\}^2},$  (16)

$$r_G = \frac{N_\lambda \operatorname{Im}\{\tilde{G}\}}{N_\lambda \operatorname{Re}\{\tilde{G}\} + |\tilde{G}|^2}$$

Here $\tilde{G} = 2G^{\uparrow\downarrow}/\sigma$, noting the above expressions are equivalent to those obtained in Ref. [10].

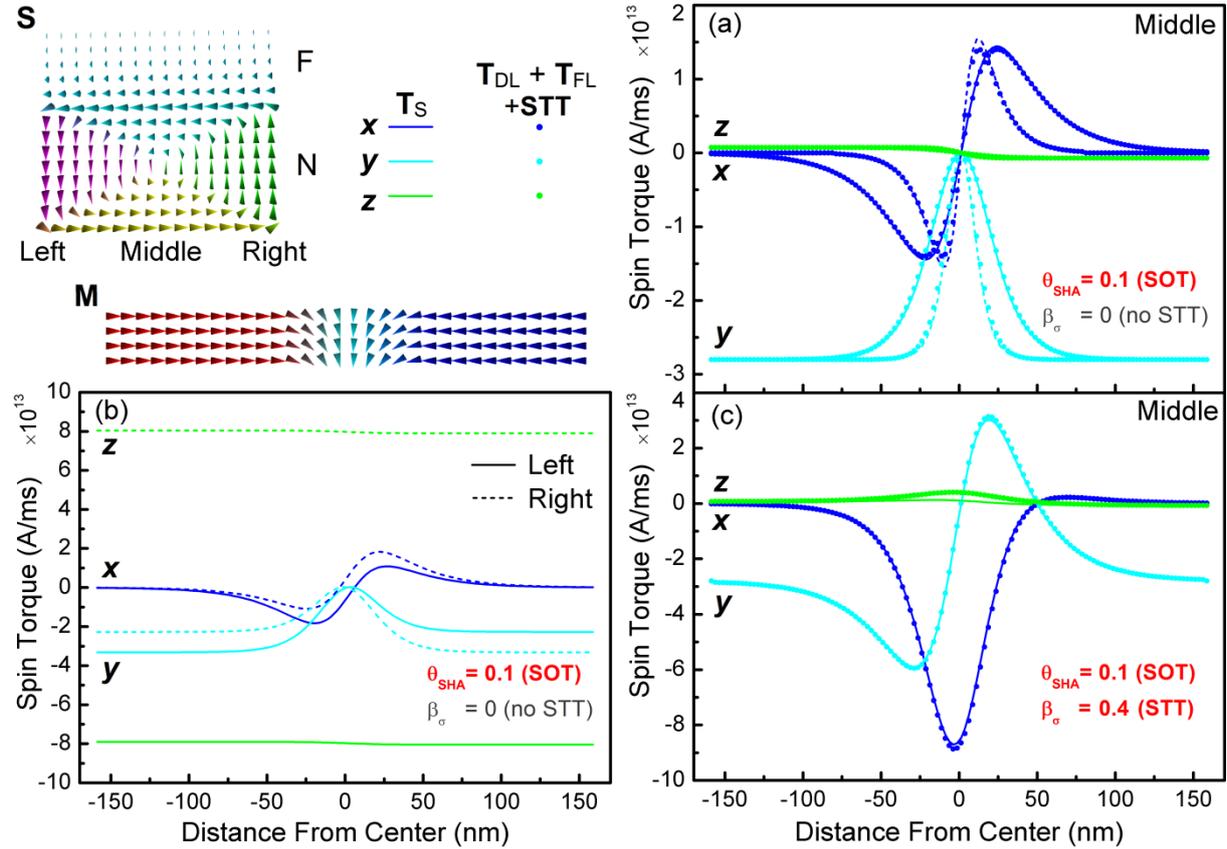

**Figure 2. Spin torques in a bilayer geometry with layers of equal width.** Spin torques calculated for a transverse domain wall in the bilayer geometry are shown for **(a)** SOT only, shown in the middle of the track for a wide (200 nm) and narrow (100 nm) domain wall width, **(b)** SOT only, shown at the left and right edges of the track, and **(c)** both SOT and STT enabled. The spin torques are shown as obtained from numerical calculations (solid lines), as well as analytical formulas (solid discs). The differences in the torques obtained at the edges and the middle of the track arise due to the non-uniformity of the spin current and spin accumulation as indicated by the rendering of **S** at the top of the figure.

Thus the two approaches result in qualitatively identical descriptions of the SHE-generated spin torques, and we note they even have identical limits: for the continuous case, taking the limit $\lambda_\varphi \to 0$ (thus the spin torque is generated purely at the interface in this limit) results in an identical torque to that obtained in the limit $\text{Re}\{G^{\uparrow\downarrow}\} \to \infty$, namely $\mathbf{T}_{FL} = 0$ and $\mathbf{T}_{DL} = -\theta_{SHA}(\mu_B/e)J_{cx}(1 - 1/\cosh(d_N/\lambda_{sf}^N))\mathbf{m} \times (\mathbf{m} \times \boldsymbol{\sigma})$. A quantitative comparison is given in Section IV.

The DL and FL torques with constant coefficients obtained above may of course be added to the LLG equation. This approach however does suffer from serious limitations. Firstly, due to the non-local nature of spin transport it can become intractable to obtain expressions for the torque coefficients in more complex geometries involving several layers, such as synthetic antiferromagnetic racetrack designs [30,31]. An even more serious limitation comes from the implicit assumption used in the above analysis, namely that the spin current incident on the N/F interface is uniform (for uniform magnetisation). For the geometry in Figure 1 this is a good approximation, however in general the charge and spin currents can be non-uniform and also the geometry thickness can vary for more complex three-dimensional devices, which would result in a complicated spatial dependence of the DL and FL torques. To reinforce this point the simple bilayer geometry is analysed again, but this time the F and N layers have the same width. Due to the rotation of the spin accumulation around the charge current direction, as seen in Figure 2, the spin current incident on the F interface is no longer uniform, resulting in a variation of the DL and FL torques across the track. Figure 2 shows results obtained using the continuous interface, however qualitatively identical results are obtained using $G^{\uparrow\downarrow}$. First, the total torque with only SHE enabled is obtained in the middle of the track for a transverse domain wall, shown in Figure 2(a). In this case equation (14) holds and a good agreement is obtained between the analytical formulas and numerical results. In Figure 2(a) the torques are calculated for two domain wall widths, 200 nm and 100 nm. A slight discrepancy arises for the narrower domain wall due to three-dimensional diffusion effects not captured by the analytical description of equation (15), however this effect is small. A much more significant discrepancy arises at the edges of the wire, see Figure 2(b), where the torques are completely different showing significant FL components.

For regions with magnetisation gradients STT also act on the magnetisation, given by Zhang and Li [38] as additional terms to the normalised LLG equation:

$$(\mathbf{u}.\nabla)\mathbf{m} - \xi \mathbf{m} \times [(\mathbf{u}.\nabla)\mathbf{m}]$$

$$\mathbf{u} = \mathbf{J}_C \frac{P}{M_S} \frac{\mu_B}{e} \frac{1}{1+\xi^2}$$

(17)

Here $P$ is the current spin polarisation, $P = (n^\uparrow - n^\downarrow)/(n^\uparrow + n^\downarrow)$ with $n^\uparrow$ and $n^\downarrow$ being the majority and minority conduction electron density of states respectively. These torques are obtained using the drift-diffusion equations in the absence of the spin dephasing length by using the valid approximation $\nabla \mathbf{S} \cong 0$, see e.g. Ref. [16], noting $\beta_\sigma = P$. When the spin dephasing length is included, the approximation $\nabla \mathbf{S} \cong 0$ is no longer valid and the non-adiabaticity parameter is modified as shown in Ref. [9]. For typical domain wall widths in Py however, equation (17) remains a good approximation with $\xi \cong \lambda_J^2 / \lambda_{sf}^2$. This is verified in Figure 2(c) where both the SOT and STT are enabled. In general the full three-dimensional treatment with the spin torque calculated using the generalised drift-diffusion equations is superior to the approach of incorporating analytical representations of the different torques in the LLG equation. The computational time is typically doubled, which is an acceptable cost given the increased accuracy, subtlety and depth of physical effects which can be modelled, as well as the convenience of a self-consistent approach to modelling spin torques.

## IV. Ferromagnetic Resonance for Spin Hall Effect Bilayers

It is well known that the DL torque in N/F bilayers modifies the linewidth obtained from FMR measurements. This is investigated here using the geometry shown in Figure 3, including contributions from SHE, Oersted field and spin pumping. Field-swept FMR peaks are simulated for bias field along the $-y$ direction and r.f. field along the $x$-axis as detailed in the Methods section − typical calculated FMR peaks are shown in Figure 3. First, the FMR peaks are simulated using the boundary conditions of equation (10), with spin pumping also included using $G^{\uparrow\downarrow} = 10^{15} + i10^{14}$ (S/m$^2$) appropriate for Pt/Py interfaces [49].

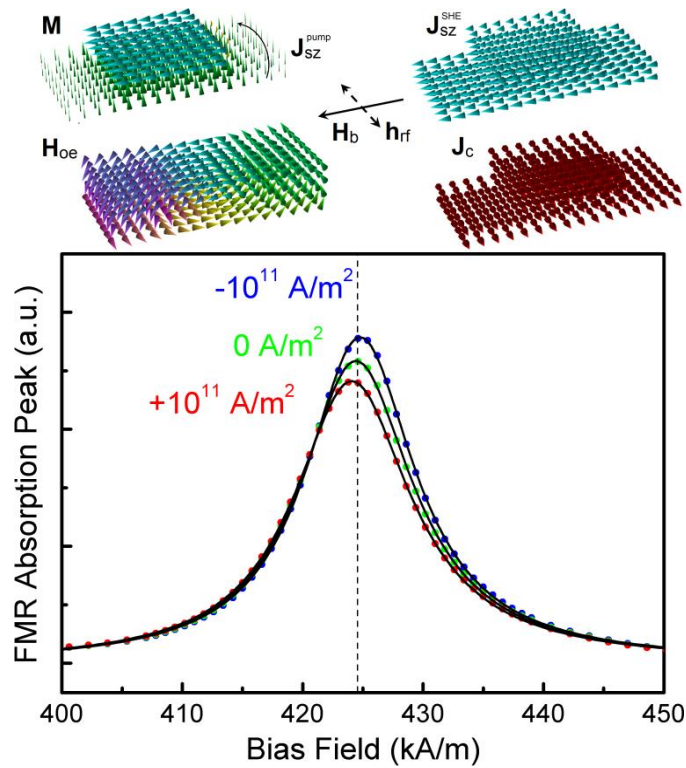

**Figure 3. Ferromagnetic resonance peaks in a spin Hall effect bilayer.** FMR peaks are shown for Pt/Py bilayers at 20 GHz, showing simulated FMR absorption peaks with fitted Lorentzian peak functions for different charge current density values in the Pt layer. The bias field and r.f. field configuration, as well as the charge and $z$ direction spin currents, including the pumped and SHE-generated spin currents, with the resulting Oersted field, are shown in the rendered images at the top.

For FMR simulations with an applied current the resonance field is shifted due to a combination of the Oersted field and the field resulting from FL component of the spin torque, namely $(rf/\gamma M_s d_F)J_{szy}$, as shown by the dotted lines in Figure 4(a), in agreement with the values extracted from the FMR peaks (solid squares). Spin pumping results in a resonance field shift independent of the current density (but dependent on the frequency) as seen in Figure 4(a), in agreement with the shift predicted due to the change in effective gyromagnetic ratio [50] – see Methods section.

The change in damping due to the DL torque is shown in Figure 4(b), calculated as a function of current density in the Pt layer for 3 different frequencies, 10, 20 and 40 GHz. Spin pumping results in a significant increase in damping of $\cong 0.0055$, where the base Gilbert damping is set to 0.01, in agreement with the expected increase for a diffusive system [50]. The damping increase due to spin pumping is constant within the fitting uncertainty both with frequency and current density. Thus by taking the difference in damping for currents with opposite direction, the resultant change $2\Delta\alpha_{SHE}$ is solely due to the SHE. As shown in Ref. [20] the change in damping is approximately inversely proportional to the r.f. frequency and directly proportional to the strength of the DL torque, given by:

$$\Delta\alpha_{SHE} \cong \frac{fJ_{szy}}{\omega M_s d_F} \tag{18}$$

For the higher frequencies a good agreement is obtained between numerical calculations and equation (18). For the lower 10 GHz frequency, close to the frequency used in Ref. [20], this relation is no longer accurate and the numerical results must be used instead.

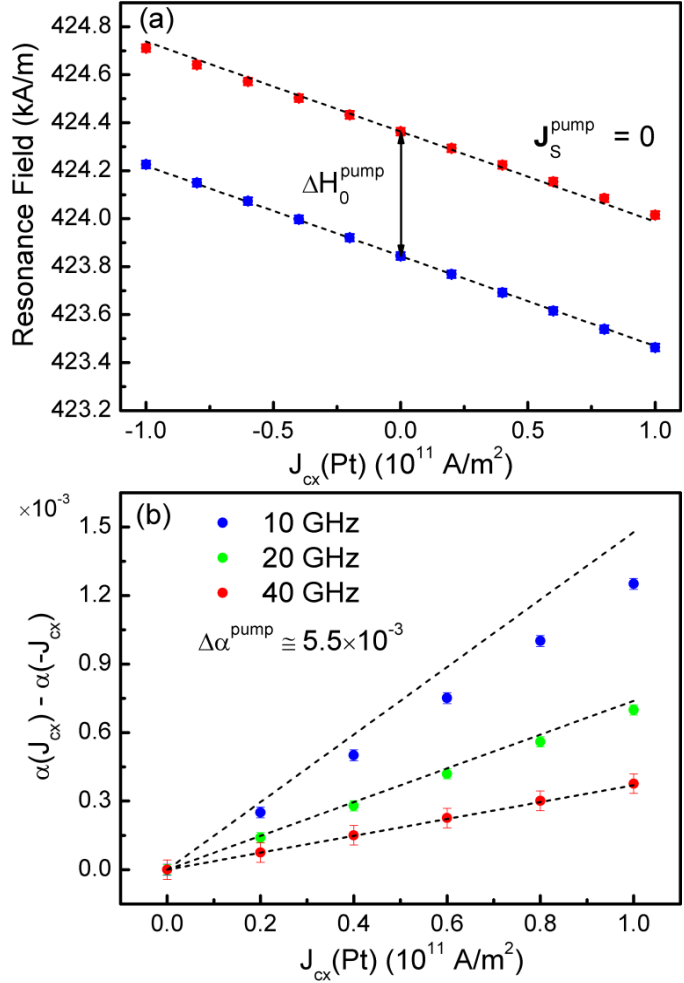

**Figure 4. Change in resonance field and damping as a function of current density.** The change in FMR peak properties as a function of current density in the Pt layer are obtained, showing **(a)** resonance field for 20 GHz frequency with the contributions of the Oersted field and FL term identified by the dotted lines, both with and without spin pumping, and **(b)** change in effective damping due to DL torque calculated for 10, 20 and 40 GHz frequency. The dotted lines are obtained from equation (18), noting the plot represents $2\Delta\alpha_{SHE}$. For both parts the error bars represent the Lorentzian peak fitting uncertainties.

The experimentally measured change in damping from Ref. [20] may be reproduced with the model introduced here by using $\theta_{SHA}$ as a fitting factor. The results are shown in Figure 5 as a function of the spin diffusion length in Pt, and for two extremes of the spin diffusion length in Py, noting $\lambda_{sdl} = \lambda_{sf}\sqrt{1-\beta_\sigma\beta_D}$ [7]. With the shorter diffusion length in Py a good agreement is obtained between the continuous interface calculations and those with a spin mixing conductance with $Re\{G^{\uparrow\downarrow}\} = 5\times10^{15}$ S/m². This value is significantly larger than the accepted value for Pt/Py interfaces, which is typically $Re\{G^{\uparrow\downarrow}\} \cong 10^{15}$ S/m² [49]. Repeating the calculations with this lower value results in $\theta_{SHA}$ in the range 0.08 – 0.1, comparable to that obtained in Ref. [20]. A quantitative agreement with the continuous interface calculations may again be obtained by using a longer diffusion length in Py of 12 nm [47] as shown in Figure 5. Note $\xi = 0.04$ is kept fixed and $\lambda_\varphi$ now takes on larger values in the range 2.5 – 4.5 nm obtained from $\lambda_\varphi = \lambda_J \sqrt{l_\perp / l_L}$ [9].

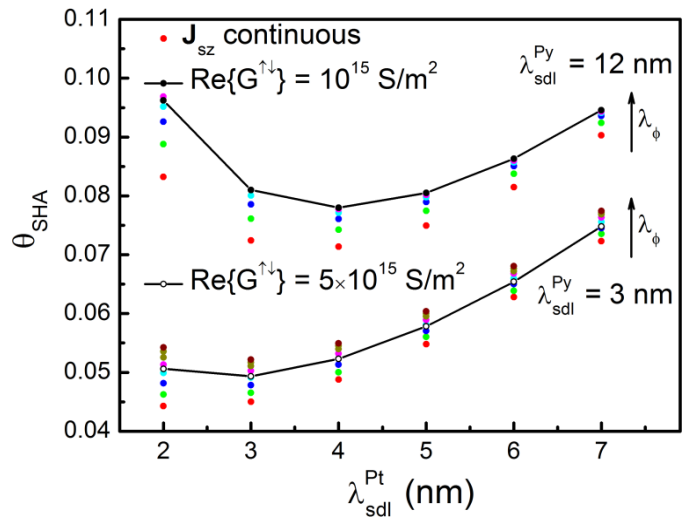

**Figure 5. Calculated spin Hall angle.** Experimental results in Ref. [20] are reproduced as a function of spin diffusion length for Pt, for both the continuous and spin mixing conductance interface calculations. For the continuous interface, results for two diffusion lengths in Py are shown as indicated, where $\lambda_\varphi$ varies in the range 0.3 – 1 nm in 0.1 nm steps for the shorter diffusion length and 2.5 – 4.5 nm in 0.5 nm steps for the longer diffusion length. In both cases the non-adiabaticity parameter is set to 0.04.

It should be noted that for both approaches the drift-diffusion model is an approximation to the stronger Boltzmann semiclassical approximation [10]. For the spin mixing conductance interface no direct dependence on the transport parameters in the F layer exists, whilst results obtained with the Boltzmann equation show the torques have a marked dependence on the F spin diffusion length [10]. On the other hand the spin torque magnitudes calculated with the continuous interface do increase with the spin diffusion length, in a qualitative agreement with the Boltzmann equation approach. However, in the continuous interface case the validity of the drift-diffusion formalism is limited to cases where the transverse spin relaxation time is greater than the momentum relaxation time. Whilst the Boltzmann equation approach is more powerful, also allowing for inclusion of current-in-plane transport effects [51], in addition to the CPP effects modelled via the present drift-diffusion approach, the computational cost is much greater [52] which may currently preclude an efficient integration within a three-dimensional model of magnetisation dynamics. For CPP transport a hybrid approach may be possible, where the spin torques are calculated using the Boltzmann equation for a static magnetisation configuration, and appropriate correction factors introduced for the spin torque magnitudes in the drift-diffusion model. In this way the advantages of a three-dimensional approach to including spin transport effects within a magnetisation dynamics simulation is maintained; the investigation of this possibility is left for further work.

V. Conclusions

A three-dimensional spin current solver based on the generalised drift-diffusion description, including the spin Hall effect and spin pumping, was implemented within a three-dimensional magnetisation dynamics formulation. This model was shown to self-consistently reproduce a number of spin torques in CPP geometries and single ferromagnetic layers. Two approaches to modelling interfaces between normal metals and ferromagnets were investigated, one based on the spin mixing conductance and the other based on continuity of spin currents. Both approaches are in qualitative agreement, showing the SHE-generated spin torques contain both field-like and damping-like components in general. A quantitative comparison between the two approaches was made by calculating the spin Hall angle in Pt from published FMR data. Whilst the spin mixing conductance approach does not directly take into account the spin diffusion length in the ferromagnet, the two approaches were shown to be in approximate agreement for published transport parameters. Finally, analytical approaches to including spin torque terms in calculations are restricted only to special cases where the spin currents incident on the metal / ferromagnet interface are uniform. In general this is not the case, as shown even for a simple bilayer geometry, and the full three-dimensional spin current solver approach is more appropriate. It is hoped this approach to modelling spin torques will lead to a better understanding of experimentally obtained spin torque-driven magnetisation dynamics.

## Appendix A – Spin Torques in CPP-GMR Stacks

It is well known that both field-like and spin-like torques act on the layers of a CPP-GMR stack [7]. These torques are of the form $a\mathbf{m}\times(\mathbf{m}\times\mathbf{m}_F) + b\mathbf{m}\times\mathbf{m}_F$, where $\mathbf{m}$ is the local magnetisation direction of the free layer, and $\mathbf{m}_F$ is the magnetisation direction of the fixed layer. For a macrospin approximation these torques may be added to the LLG equation with appropriate values for the coupling constants $a$ and $b$ [53]. In general however these coefficients depend on the spin accumulation and have a spatial dependence. Moreover the spin accumulation is important in understanding the magneto-resistance of the CPP-GMR stack [54]. For completeness the model implemented here is tested in a simple spin valve, showing simultaneous reproduction of both the spin torque switching effect, as well as the magneto-resistance effect. The full micromagnetics model is used, including demagnetising and direct exchange contributions, coupled to the three-dimensional spin current solver. The CPP-GMR stack consists of the layering N (59 nm) / F (5 nm) / N (2 nm) / F (3 nm) / N (59 nm). The thicker F layer is the fixed magnetic layer where the magnetisation is kept fixed along the $x$ direction, and the thinner F layer is the free magnetic layer. This is shown in Figure A1, where two extreme configurations are distinguished: the anti-parallel configuration (AP) where the magneto-resistance is the highest, and the parallel configuration (P) where the magneto-resistance is the lowest. The stack is elliptical in shape with 160 nm × 40 nm dimensions, and the electrodes are placed at the $z$-axis ends of the structure. Stair-step boundary corrections are applied to the elliptical shape to correct for the finite difference artefacts on the demagnetising field [42,55]. The outer N leads are purposely extended (for simplicity here they are simply extended along the $z$-axis, but more complicated contact geometries are possible) to allow the spin accumulation to decay to zero – it is important to include in the model enough of the contacting electrical leads since only then can the boundary condition $(\nabla \mathbf{S}).\mathbf{n} = 0$ be applied correctly. The resulting $x$ components of the spin accumulation for the AP and P states are shown in Figure A1b. The same material parameters for the N and F layers used in the main text are applied here using the spin mixing conductance interface approach (but with $\theta_{SHA} = 0$); additionally $\beta_D = 0.9$ [47]. The results are shown in Figure A1, where the magnetisation of the free layer is switched from the AP to the P state using a charge current density along the $z$ direction of $-10^{12}$ A/m$^2$ (electrons flow from the fixed to the free layer), and back to the AP state using a charge current density of $+10^{12}$ A/m$^2$. For simplicity the layers are not surface-exchange coupled, but interact only through the demagnetising field and the spin torque. As expected the resistance switches from a high state (AP) to a low state (P), and back to the original state. Since the demagnetising field preferentially acts to keep the layers in the AP state, the switching process is slower from the AP to the P state than vice-versa, however once the P state is achieved the shape anisotropy of the ellipse stabilises this configuration.

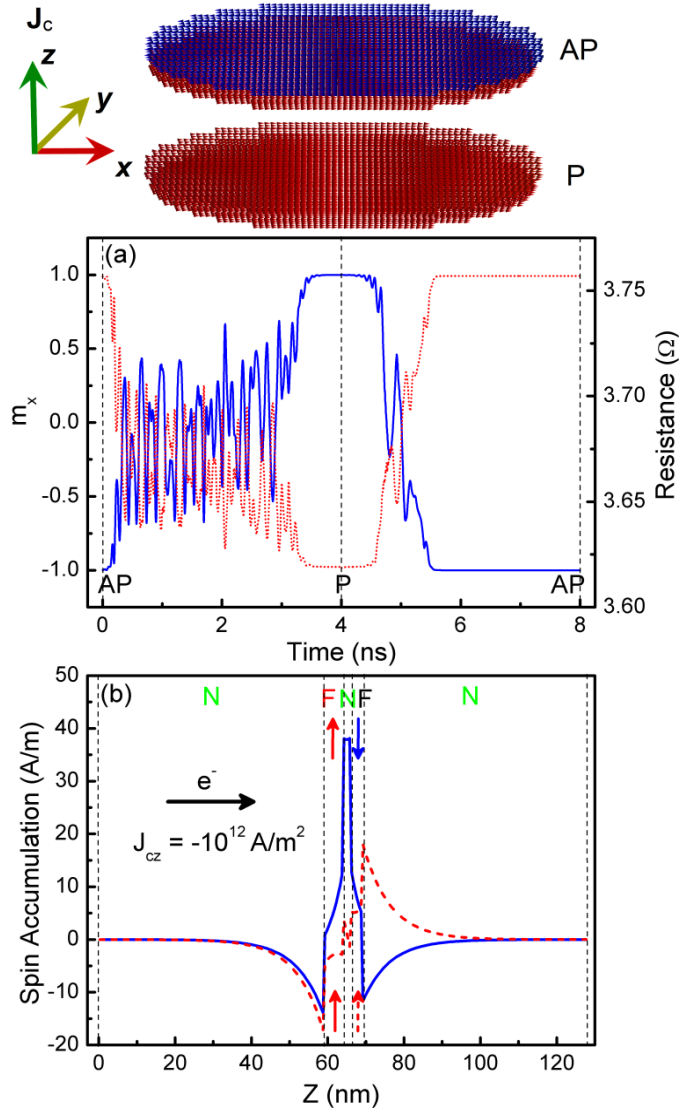

**Figure A1. Current-induced magnetisation switching in a spin valve.** Spin torque switching in an elliptically shaped CPP-GMR stack, showing **(a)** normalised *x* component of magnetisation (solid line) and stack resistance (dotted line) showing switching from the AP state (using $J_{cz} = -10^{12}$ A/m$^2$) to the P state and back (using $J_{cz} = +10^{12}$ A/m$^2$) to the AP state, **(b)** *x* component of spin accumulation for the AP and P configurations.

Appendix B – Methods

All simulations were done using the multi-physics micromagnetics-oriented software Boris [42,43] written by the author. The software is written mainly in C++ with all computational routines available for both CPU and GPU computations using the CUDA C framework. For small problem sizes but highly repetitive computations, such as the FMR simulations, CPU routines are more efficient. To optimise the computational speed the most expensive routines have been written directly in assembly language using the SIMD AVX instruction set. In particular taking advantage of the larger AVX registers allowing 2 FFTs to be computed simultaneously on each processor core, custom interleaved FFT routines have been implemented using the radix 4 algorithm (this was found to be more efficient than the more common split radix algorithm). For larger problem sizes, such as the CPP-GMR stack, GPU routines are increasingly more efficient due to massive parallelisation. For more complex simulations and simulation chains, such as the FMR simulations, instead of console-based user control the compiled program is controlled using local or remote Python scripts with a communication protocol implemented over network sockets.

For the FMR simulations, for each bias field value the magnetisation precession is allowed to reach steady state before extracting the oscillation amplitude. At the end of the bias field sequence the Lorentzian peak function $F(x) = y_0 + S [\Delta H + A (x-H_0)] / [4(x-H_0)^2 + \Delta H^2]$ – this formula contains both symmetric and asymmetric components, however the fitted formula contains virtually only the symmetric component ($A \cong 0$) as expected – is automatically fitted using the Levenberg-Marquardt algorithm. Typical fitted FMR peaks obtained at 20 GHz frequency are shown in Figure 3, where the quoted charge current density is the average value in the N layer. To obtain a peak representative of the FMR power absorption the oscillation amplitude is squared: the resulting peak is described very well by the Lorentzian peak function from which the damping value can be extracted as $\alpha = \gamma \Delta H / 2\omega$ [56], where $\Delta H$ is the full-width half-maximum linewidth and $\omega$ is the angular frequency. The zero-current FMR peaks have a resonance field $H_0$ close to that predicted by the Kittel formula [56] $\omega = \gamma ([H_0 + (N_y - N_x)M_s][H_0 + (N_z - N_x)M_s])^{0.5}$, where $N_x = N_y = 0.113$ and $N_z = 0.774$ are demagnetising factors calculated for the Py rectangle; note, an exact agreement with this formula cannot be expected since it only strictly applies to ellipsoidal shapes. Spin pumping results in a change in the effective gyromagnetic ratio and effective damping. For an ideal spin sink these can be expressed in terms of the spin mixing conductance, see equations (59) and (60) in Ref. [50]. For example at 20 GHz the predicted reduction in resonance field is ~430 A/m, comparable to the value obtained from simulations of ~480 A/m. Similarly the predicted increase in damping for an ideal spin sink is 0.0079, consistent with the lower value obtained from simulations of 0.0055, expected for a non-ideal diffusive spin sink.

The LLG equation was solved using the 4$^{th}$ order Runge-Kutta method with fixed time-step of 0.1 ps. Equation (6) is solved using a custom FFT-based Poisson solver. Equation (8) is solved using an alternating direction scheme [44]. The cellsizes used are 5 nm for the LLG equation (refined to 1 nm in the z direction for the CPP-GMR stack) with the spin current solver sub-mesh refined along the *z*-axis to 0.125 nm.